\documentclass[twocolumn,showpacs, showkeys,nofootinbib,preprintnumbers,amsmath,amssymb,superscriptaddress]{revtex4-1}

\usepackage{graphicx} 
\usepackage{dcolumn} 
\usepackage{bm} 
\usepackage{color} 
\usepackage{slashed}
\usepackage{dsfont}

\usepackage{hyperref}

\newcommand{\mrm}[1]{\mathrm{#1}}
\newcommand{\mc}[1]{\mathcal{#1}}
\newcommand{\pd}{\partial}
\newcommand{\blr}[1]{\left(#1\right)}
\newcommand{\be}{\begin{equation}}
\newcommand{\ee}{\end{equation}}
\newcommand{\bea}{\begin{eqnarray}}
\newcommand{\eea}{\end{eqnarray}}
\newcommand{\bse}{\begin{subequations}}
\newcommand{\ese}{\end{subequations}}
\newcommand{\su}{\sigma_{\!u}}
\newcommand{\sd}{\sigma_{\!d}}
\newcommand{\sy}{\sigma_{\!s}}
\def\Phib{\bar{\Phi}}
\def\qqb{{q\bar{q}}}
\def\eq#1{(\ref{#1})}
\def\Eq#1{Eq.~(\ref{#1})}
\newcommand{\Tr}{\mrm{Tr}}

\graphicspath{{./figures/}}

\definecolor{orange}{rgb}{1,0.5,0}

\begin{document}

\title{Thermodynamics of $(2+1)$-flavor strongly interacting matter at nonzero isospin}

\author{Rainer Stiele}\email{r.stiele@thphys.uni-heidelberg.de}
\affiliation{Institut f\"{u}r Theoretische Physik, Universit\"at Heidelberg, Philosophenweg 16, D-69120 Heidelberg, Germany}

\author{Eduardo S.~Fraga}\email{fraga@if.ufrj.br}
\affiliation{Institut f\"{u}r Theoretische Physik, Goethe-Universit\"at Frankfurt, Max-von-Laue-Stra\ss e 1, D-60438 Frankfurt am Main, Germany}
\affiliation{Instituto de F\'\i sica, Universidade Federal do Rio de Janeiro, Caixa Postal 68528, Rio de Janeiro, RJ 21945-970, Brazil}

\author{J\"urgen Schaffner-Bielich}\email{schaffner@astro.uni-frankfurt.de}
\affiliation{Institut f\"{u}r Theoretische Physik, Goethe-Universit\"at Frankfurt, Max-von-Laue-Stra\ss e 1, D-60438 Frankfurt am Main, Germany}

\begin{abstract}
We investigate the phase structure of strongly interacting matter at non-vanishing isospin before the onset of pion condensation in the framework of the unquenched Polyakov--Quark-Meson model with $2+1$ quark flavors. 
We show results for the order parameters and all relevant thermodynamic quantities. In particular, we obtain a moderate change of the pressure with isospin at vanishing baryon chemical potential, whereas the chiral condensate decreases more appreciably.
We compare the effective model to recent lattice data for the decrease of the pseudo-critical temperature with the isospin chemical potential. We also demonstrate the major role played by the value of the pion mass in the curvature of the transition line, and the need for lattice results with a physical pion mass. Limitations of the model at nonzero chemical potential are also discussed.
\end{abstract}

\pacs{25.75.Nq,11.30.Rd,12.38.Aw,11.10.Wx}

\keywords{quark deconfinement, chiral symmetry, isospin}

\maketitle

\section{Introduction}
\label{sec:Intro}

The thermodynamics of strongly interacting matter under extreme conditions plays a major role in the understanding of the physical scenario shortly after the Big Bang \cite{Boyanovsky:2006bf,Schettler:2010wi}, in the outcome of high-energy heavy ion collisions \cite{Marty:2012vs}, in the mechanism of supernovae explosions \cite{Janka:2012wk}, and in the structure of compact stars \cite{Weber:2004kj,SchaffnerBielich:2010am}. The possibility to probe such large temperatures and densities in current experiments at LHC-CERN and RHIC-BNL, and especially in future experiments at FAIR-GSI, calls for a detailed study of the transition to the chirally symmetric quark-gluon plasma phase and  the properties of this new extraordinary state of matter \cite{Muller:2006ee,Muller:2012zq,Bleicher:2011sd}.

In all systems mentioned above matter does not consist of equal amounts of protons and neutrons, i.e. one has a non-vanishing isospin density. Using $Au$ or $Pb$ beams in heavy ion collisions, the proton to neutron ratio is  $\sim$ 2/3. In astrophysical environments the initial proton fraction in supernovae is $0.4$, reduces to $0.2$ and finally reaches values of less than $0.1$ in cold neutron stars. In the universe a large asymmetry in the lepton sector is allowed ($-0.38<\mu_\nu/T<0.02$) \cite{Schwarz:2012yw}, which can shift the equilibrium conditions  at the cosmological QCD transition \cite{Schwarz:2009ii}. Hence, in the description of the thermodynamics in all these scenarios of nature isospin should not be overlooked.

In this work we investigate the phase structure of strongly interacting matter at non-vanishing isospin before the onset of pion condensation in the framework of the Polyakov--Quark-Meson model with $2+1$ quark flavors. First we compute the chiral condensate, the pressure and the trace anomaly at vanishing quark densities, reproducing well-established lattice data. Then we compute all relevant thermodynamic quantities and the phase diagram for a nonzero isospin chemical potential. For vanishing baryon chemical potential, our results can be compared directly to lattice data since this case is free of the Sign Problem \cite{deForcrand:2006pv,DeTar:2009ef}.

Effective chiral models, combined with some version of the Polyakov loop potential, usually have their parameters adjusted to provide a good description of lattice data at zero density. Testing effective models built in this fashion against lattice data at nonzero isospin is crucial to understand whether their extended versions provide qualitative and quantitative accurate descriptions of the phase structure of strong interactions.

To date, most of the calculations at non-vanishing isospin were done with only two quark flavors, neglecting strange quarks as relevant degrees of freedom at the energy scale of the chiral and deconfinement transition \cite{Kogut:2004zg,deForcrand:2007uz,Cea:2012ev, Fraga:2008be,Palhares:2008px,Toublan:2003tt,Frank:2003ve,Barducci:2004tt,He:2005nk,Ebert:2005cs,Ebert:2005wr,Andersen:2007qv,Abuki:2008wm,Kang:2013bea,Kamikado:2012bt,Sasaki:2010jz,Cavagnoli:2013mba,Ueda:2013sia}.
Furthermore, lattice calculations at nonzero isospin were performed so far only with unphysical heavy quark masses \cite{Kogut:2004zg,deForcrand:2007uz,Cea:2012ev}. The impact of the quark mass on the deconfining critical temperature at nonzero isospin was investigated in Refs.~\cite{Fraga:2008be,Palhares:2008px}, using a framework that combines chiral perturbation theory to describe the low-energy sector with the phenomenological fuzzy bag model for high energies, showing that quark masses play a relevant role. Previously, there have also been investigations with the hadron resonance gas model \cite{Toublan:2004ks}, the $O(2N)$-symmetric $\phi^4$-theory \cite{Andersen:2006ys}, the Nambu--Jona-Lasinio model \cite{Toublan:2003tt,Frank:2003ve,Barducci:2004tt,He:2005nk,Ebert:2005cs,Ebert:2005wr,Andersen:2007qv,Abuki:2008wm,Xia:2013caa,Kang:2013bea}, the Quark-Meson model \cite{Kamikado:2012bt}, and its Polyakov-loop extended versions \cite{Sasaki:2010jz,Cavagnoli:2013mba,Ueda:2013sia}.

One of the aspects in the extension of the phase diagram to nonzero isospin is the arising of a new phase. Charged pions couple to the isospin chemical potential and at $\mu_I=\mu_u-\mu_d=m_\pi$ there is the onset of pion condensation \cite{Son:2000xc}. The running of the pion mass in the medium shifts the appearance of pion condensation to larger temperatures and densities.
Depending on the analyzed region of isospin, temperature and baryon chemical potential, pion condensation must be taken into account \cite{Kogut:2004zg,deForcrand:2007uz,Fraga:2008be,Palhares:2008px,Andersen:2006ys,Barducci:2004tt,He:2005nk,Ebert:2005cs,Ebert:2005wr,Andersen:2007qv,Abuki:2008wm,Kamikado:2012bt,Sasaki:2010jz,Xia:2013caa,Kang:2013bea} or not \cite{Cea:2012ev,Toublan:2004ks,Toublan:2003tt,Ueda:2013sia,Frank:2003ve}.

In contrast to what is considered in this work, the two-flavor renormalization group (RG) improved Quark-Meson model was so far only applied to analyze the pion condensate phase \cite{Kamikado:2012bt} and investigations with the PNJL model only considered two quark flavors and applied the simpler pure gauge Polyakov-loop potential \cite{Sasaki:2010jz,Cavagnoli:2013mba,Ueda:2013sia}.

We restrict our analysis to moderate isospin chemical potential values, before the onset of pion condensation. To describe strongly interacting matter we adopt the framework of the Polyakov-Quark-Meson model \cite{Meisinger:1995ih,Pisarski:2000eq,Fukushima:2003fw,Megias:2004hj,Ratti:2005jh,Schaefer:2007pw,Schaefer:2009ui,Herbst:2010rf,Schaefer:2011ex,Gupta:2011ez,Chatterjee:2011jd,Mintz:2012aua, Mintz:2012mz,Herbst:2013ail,Haas:2013qwp,Stiele:2013gra,Herbst:2013ufa} that we enhance by applying the unquenched Polyakov-loop potential \cite{Haas:2013qwp,Stiele:2013gra}. We build its extension to nonzero isospin in Sec.~\ref{sec:PQMmodel}. In Sec.~\ref{sec:RD} we discuss our results on the evolution of the order parameters and thermodynamics with increasing isospin. We compare the decrease of the pseudo-critical temperature with isospin chemical potential to recent lattice data to test the model in its extension to nonzero isospin density. Moreover, we point out the impact of the pion mass on the curvature of the transition line.

\section{Theoretical framework}
\label{sec:PQMmodel}

We perform our investigation within the framework of a low energy effective model that includes important aspects of QCD: chiral symmetry breaking and (partial) confinement (in the gluonic sector).
These properties are contained in the effective Lagrangian of the theory 
\bea
	\mc{L} &=& \bar{q} \blr{i\slashed{D} - g \,\phi_5 + \gamma_0 \,\mu_f} q + \Tr\blr{\pd_\mu \phi^\dagger \,\pd^\mu\phi} -\nonumber\\
	&&- \,m^2\, \Tr \blr{\phi^\dagger\phi} - \lambda_1 \left[ \Tr\blr{\phi^\dagger\phi} \right]^2 - \lambda_2\, \Tr \blr{\phi^\dagger\phi}^2 + \nonumber\\
	&&+\, c \blr{\det\phi + \det\phi^\dagger} + \Tr \left[ H \blr{\phi + \phi^\dagger} \right] - \nonumber\\
	&& -\, \mc{U} \blr{\Phi,\Phib}\;,
	\label{eq:Lagrangian}
\eea
where $\phi$ and $\phi_5$ are 3$\times$3 matrices that combine scalar and pseudoscalar meson fields.
All the contributions to the Lagrangian are discussed in the following as well as in detail e.g.~in Refs.~\cite{Lenaghan:2000ey,Kovacs:2007sy,Schaefer:2009ui,Schaefer:2008hk,Schaefer:2011ex,Chatterjee:2011jd}.
We enhance it by the application of the unquenched Polyakov-loop potential $\mc{U}_\mrm{glue}$ \cite{Haas:2013qwp,Stiele:2013gra}. To describe the creation of constituent quark masses by spontaneous breaking of chiral symmetry we use a $(2+1)$-flavor Quark-Meson model \cite{Lenaghan:2000ey,Schaefer:2008hk}. To include isospin effects, we generalize the self-interaction potential of the meson fields to distinguish between the up and down quark condensates \cite{Stiele:2012bda,Beisitzer:2013}
\begin{multline}
	\hskip-2exU\blr{\su,\sd,\sy} = \frac{\lambda_1}{4} \left[ \blr{\frac{\su^2+\sd^2}{2}}^2 + \sy^4 + \blr{\su^2 + \sd^2} \sy^2 \right] +\\
	+ \frac{\lambda_2}{4} \blr{\frac{\su^4+\sd^4}{4}+\sy^4} - \frac{c}{2\sqrt{2}} \su \sdÊ\sy +\\
	+ \frac{m^2}{2} \blr{\frac{\su^2+\sd^2}{2}+\sy^2} - \frac{h_{ud}}{2} \blr{\su+\sd} - h_{s} \sy \;.
	\label{eq:LSpotential}
\end{multline}
This potential contains spontaneous and explicit breaking of chiral symmetry. In accord with the Vafa-Witten theorem \cite{Vafa:1983tf} isospin symmetry of the vacuum is not broken in QCD and therefore we consider only one single explicit symmetry breaking term for the up-and-down quark sector $h_{ud}$.

An effective description of confinement can be implemented by the Polyakov loop as an order parameter of center symmetry \cite{Polyakov:1978vu}. Further details can be found in Ref.~\cite{Mintz:2012mz}. For the Polyakov-loop potential, one should choose a functional form that is invariant under center symmetry transformations \cite{Pisarski:2000eq, Scavenius:2002ru,Fukushima:2003fw, Ratti:2005jh, Roessner:2006xn}. Different parametrizations of the potential are available, one of which is the polynomial form of Ref.~\cite{Ratti:2005jh}
\be
	\frac{\mc{U}\blr{\Phi,\Phib;\,t,\mu_f}}{T^{4}} = -\frac{b_2\blr{t}}{2} \Phib\Phi - \frac{b_3}{6} \blr{\Phi^3+\Phib^3}+\frac{b_4}{4} \blr{\Phib\Phi}^2 \;,
	\label{eq:Polyakov_potential_polynomial}
\ee
with the temperature-dependent coefficient $b_2$ defined as
\be
	b_2\blr{t} = a_0 + \frac{a_1}{1+t} + \frac{a_2}{\blr{1+t}^2} + \frac{a_3}{\blr{1+t}^3}\;.
	\label{eq:Polyakov_potential_coefficients}
\ee
Here $t = \blr{T-T_0}/T_0$ is the reduced temperature with the critical temperature of the Polyakov-loop potential given by $T_0$.
\begin{table}
	\caption{Parameters of the Polyakov loop potential of Ref.~\cite{Scavenius:2002ru} converted in Ref.~\cite{Mintz:2012mz} to the definition of the coefficients in Eqs.~\eq{eq:Polyakov_potential_polynomial} and \eq{eq:Polyakov_potential_coefficients}.}
	\begin{ruledtabular}
		\begin{tabular}{cccccc}
			$a_0$ & $a_1$ & $a_2$ & $a_3$ & $b_3$   & $b_4$ \\ \hline
			1.53 & 0.96  & -2.3  & -2.85 & 13.34   & 14.88 \\
		\end{tabular}
	\end{ruledtabular}
	\label{tab:Ploop_pot_params}
\end{table}
The parameters of the potential are chosen such that it reproduces the temperature dependence of the Polyakov-loop expectation value and the thermodynamics of pure gauge theory as obtained in lattice calculations \cite{Scavenius:2002ru,Ratti:2005jh,Roessner:2006xn}. We use the parameter set of Ref.~\cite{Scavenius:2002ru}, which we summarize in Table \ref{tab:Ploop_pot_params}.\\
To convert this Yang-Mills potential for the Polyakov loop to the glue potential of full QCD we use the relation
\be
	t_{YM} (t_{glue}) \approx 0.57\, t_{glue}
	\label{eq:tYMtglue}
\ee
that connects the temperature scales of both theories \cite{Haas:2013qwp,Stiele:2013gra}. This rescaling accounts for the back-reaction of quarks on the gluon sector at zero quark density \cite{Braun:2009gm,Pawlowski:2010ht}.
Furthermore, we include the running of the critical temperature of the Polyakov-loop potential with the quark densities. Therefore, we generalize the description presented in Refs.~\cite{Schaefer:2007pw,Herbst:2013ail} to include different chemical potentials for each quark flavor,
\be
	T_0\blr{\mu_f} = T_\tau\, e^{ -1/\blr{\alpha_0 b\blr{\mu_f}}} \;,
\ee
with
\be
	b\blr{\mu_f} = \bar{b}_0 - \frac{16}{\pi}\sum_{N_f}\frac{\mu_f^2}{T_\tau^2}\frac{\bar{T}_0^2}{\bar{T}_0^2+m_f^2} \;.
	\label{eq:T0muf}
\ee
Here, $T_\tau$ is the UV scale that is fixed to the mass of the $\tau$-lepton, $m_\tau=1777\,\mrm{GeV}$ which gives a coupling $\alpha_0\simeq0.303$ consistent with observations \cite{Beringer:1900zz}.
The parameter $\bar{b}_0$ can be adjusted to consider a dependence on the number of quark flavors. We choose $\bar{b}_0=11N_c/{6\pi}$ such that $\bar{T}_0=270\,\mrm{MeV}$.
Here $m_f$ stands for the current quark masses and we adopt $m_s=95\,\mrm{MeV}$ \cite{Beringer:1900zz}.

\begin{figure*}
 	\includegraphics[width=.327\textwidth]{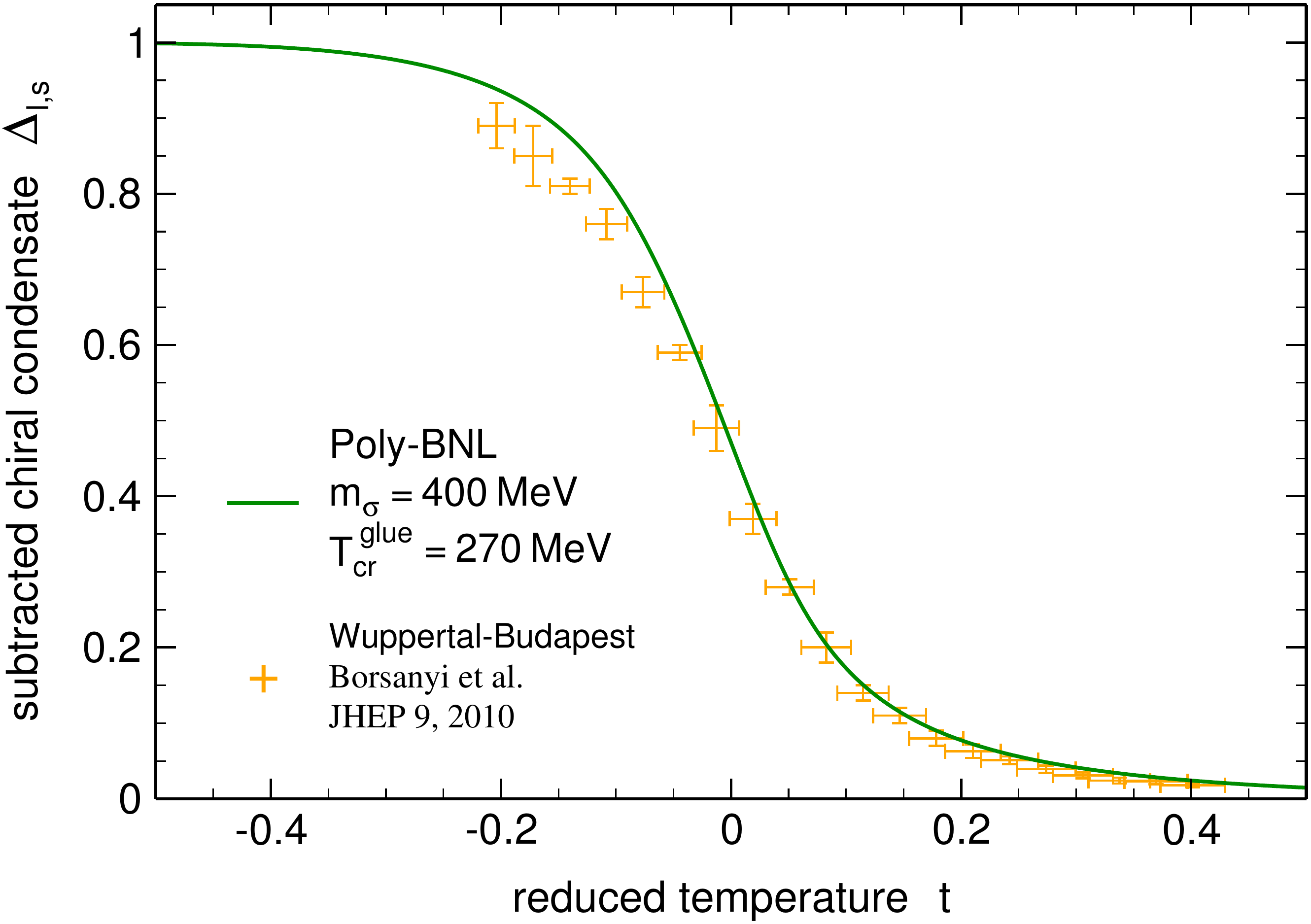}
	\hfill
	\includegraphics[width=.327\textwidth]{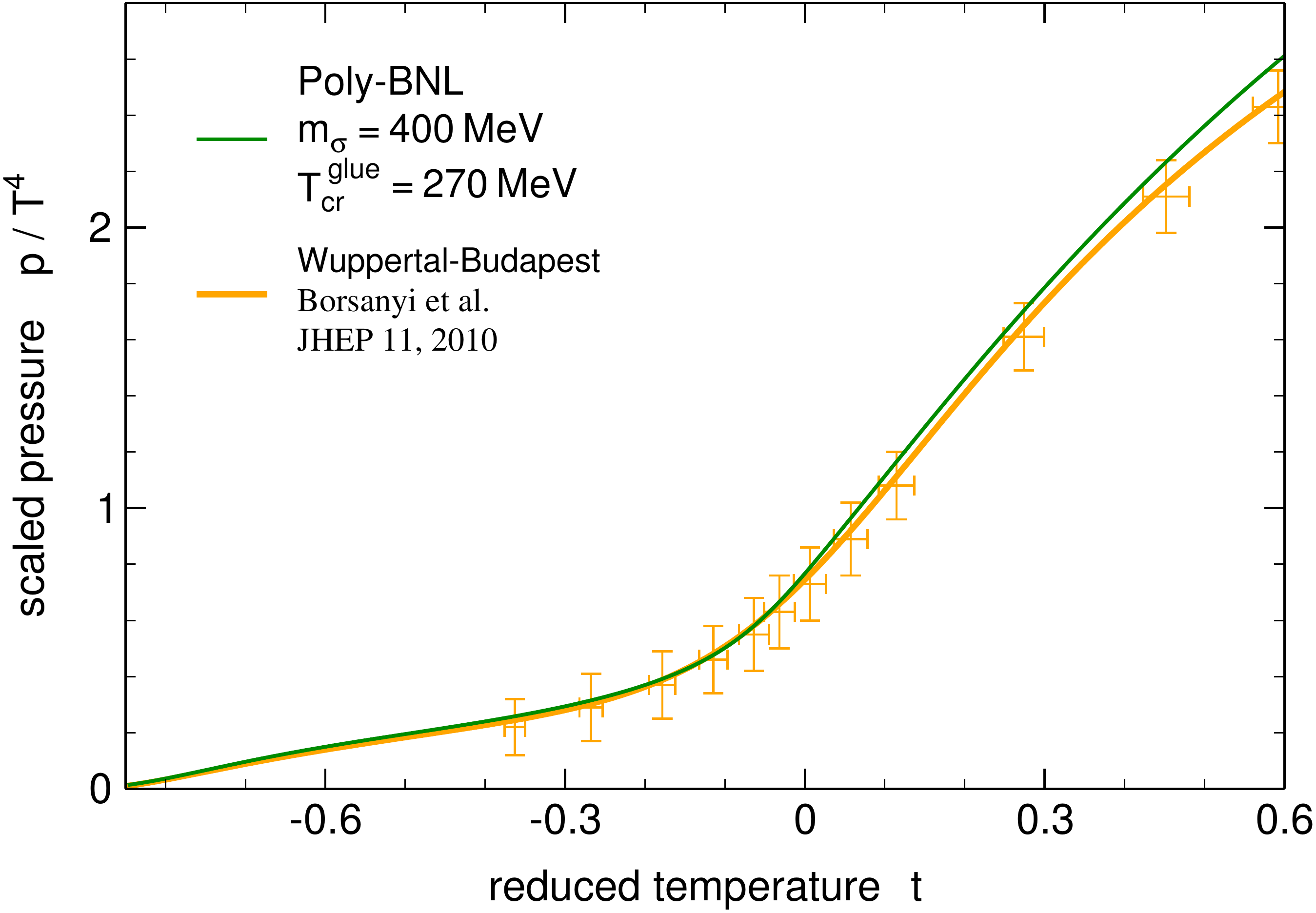}
	\hfill
	\includegraphics[width=.327\textwidth]{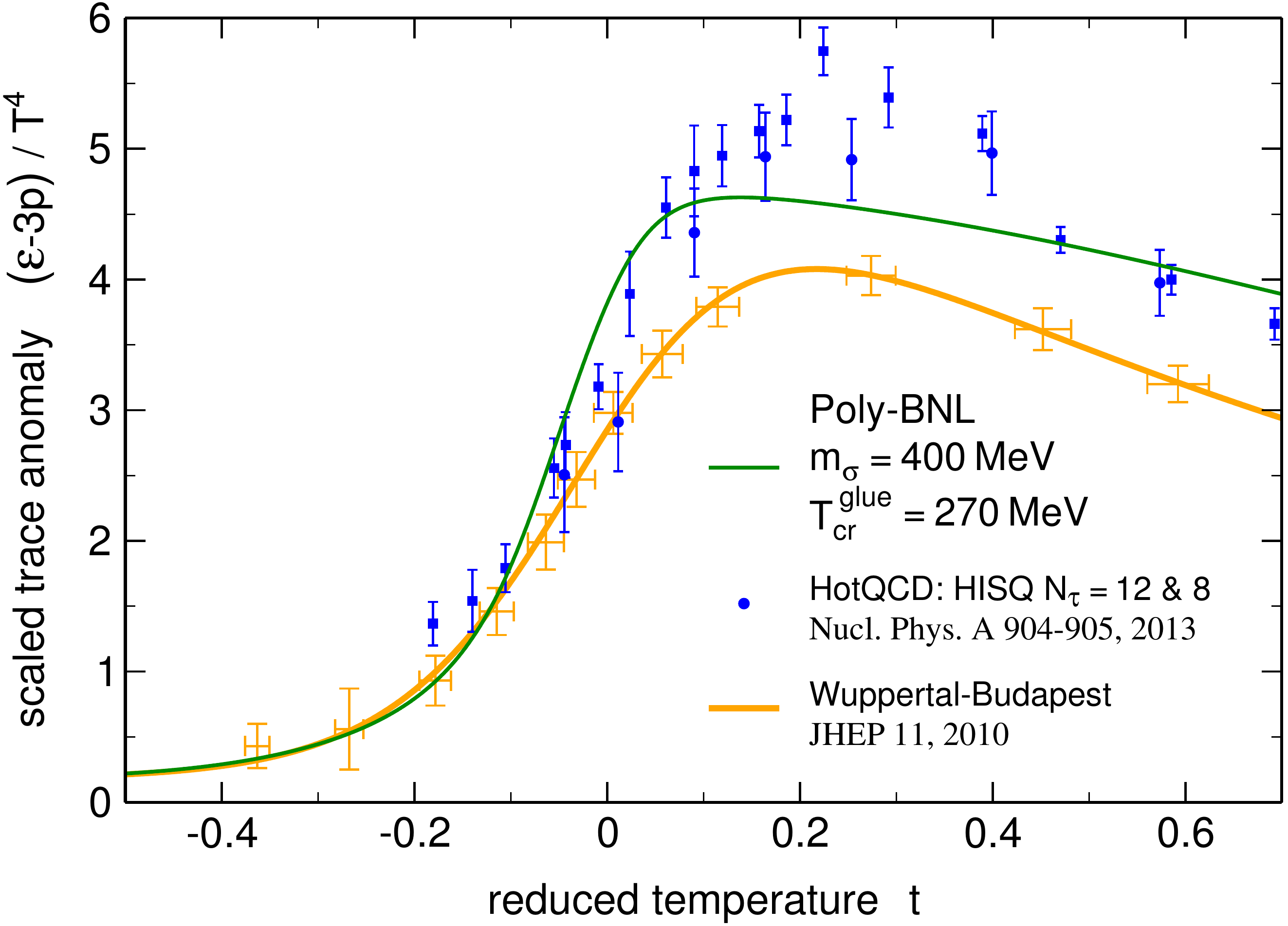}
	\caption{Subtracted chiral condensate and normalized pressure and trace anomaly as functions of the temperature at vanishing quark densities. We compare our results (full green lines) to the lattice calculations of Refs.~\cite{Borsanyi:2010bp,Borsanyi:2010cj,Bazavov:2012bp}. We use the polynomial parametrization of the Polyakov-loop potential with the parameter set of Ref.~\cite{Scavenius:2002ru} (abbreviated by Poly-BNL) with $T_\mrm{cr}^\mrm{glue}=270\,\mrm{MeV}$ as its transition scale and a $\sigma$-meson mass of $400\,\mrm{MeV}$.}
	\label{fig:muI_0}
\end{figure*}

The thermal fluctuation contribution from quarks and anti-quarks, which comes from the thermal fermionic determinant,
\begin{multline}
	\label{eq:qqbPotential}
	\Omega_\qqb^{th}\blr{\su,\sd,\sy,\Phi,\Phib;T,\mu_f}=-2T\sum_{f={u,d,s}}\int\frac{d^3p}{(2\pi)^3}\times\\
    	\times \left\{\ln\left[1 + 3\blr{\Phi + \Phib e^{-\blr{E_f-\mu_f}/T}}\times  \right.\right. \\
    	\left.\times e^{-\blr{E_f-\mu_f}/T} + e^{-3\blr{E_f-\mu_f}/T}\right] + \\
	+ \ln\left[1 + 3\blr{\Phib + \Phi e^{-\blr{E_f+\mu_f}/T}}\times \right. \\
	\left.\left.\times e^{-\blr{E_f+\mu_f}/T} +
            e^{-3\blr{E_f+\mu_f}/T}\right]\right\}\qquad
\end{multline}
includes the coupling to and between the Polyakov-loop variables and meson fields.
The dispersion relation of the quarks is given by $E_{f} = \sqrt{k^2+m_f^2}$, with the individual constituent quark masses
\be
	\label{eq:quarkmasses_coupling}
	m_u = \frac{g}{2} \su\;, \qquad m_d = \frac{g}{2} \sd \qquad \text{and} \qquad m_s = \frac{g}{\sqrt{2}} \sy\;.
\ee
We fix the Yukawa coupling $g$ between quarks and mesons by choosing $m_{l} = 300\,\mrm{MeV}$ as the constituent mass of the light quarks.

As another step beyond the usual mean-field analysis we take into account the fermionic vacuum loop contribution \cite{Mocsy:2004ab,Palhares:2008yq,Fraga:2009pi,Boomsma:2009eh,Skokov:2010sf} to the Polyakov--Quark-Meson model \cite{Schaefer:2011ex,Gupta:2011ez,Chatterjee:2011jd}
\begin{multline}
	\Omega_\qqb^{vac}\blr{\su,\sd,\sy} = -2N_c\sum_{f={u,d,s}}\int\frac{d^3p}{(2\pi)^3}E_f\\ = -\frac{N_c}{8\pi^2} \sum_{f={u,d,s}} m_f^4 \ln\blr{\frac{m_f}{\Lambda}} \;.
	\label{eq:QuarkVac}
\end{multline}
We include this contribution in the adjustment of the parameters of \Eq{eq:LSpotential} to the physical meson masses and pion and decay constants in the vacuum that are given in Table \ref{tab:chiral_pot_constants}. The dependence of \Eq{eq:QuarkVac} on the regularization scale $\Lambda$ cancels neatly with that of these parameters as is shown in detail in Ref.~\cite{Chatterjee:2011jd}.
\begin{table}
	\caption{Values of constants to which the parameters of the
          mesonic potential are adjusted, according to
          Ref.~\cite{Beringer:1900zz}.}
	\begin{ruledtabular}
		\begin{tabular}{l|ccccccc}
			Constant & $f_\pi$ & $f_K$ & $m_\pi$ & $m_K$ & $m_\eta$ & $m_{\eta'}$ & $m_\sigma$ \\ \hline
			Value [MeV] & 92 & 110 & 138 & 495 & 548 & 958 & 400 \\
		\end{tabular}
	\end{ruledtabular}
	\label{tab:chiral_pot_constants}
\end{table}

All contributions discussed above add up to the grand canonical potential
\begin{multline}
	\Omega\blr{\su,\sd,\sy,\Phi,\Phib;\,T,\mu_f} \\= U\blr{\su,\sd,\sy} + \Omega_\qqb^{vac}\blr{\su,\sd,\sy} +\\+ \mc{U}\blr{\Phi,\Phib;\,T,\mu_f} + \Omega_\qqb^{th}\blr{\su,\sd,\sy,\Phi,\Phib;\,T, \mu_f}
	\label{eq:grand_canon_pot}
\end{multline}
which is the central quantity to determine the expectation values of the order parameters as well as the thermodynamic quantities at given temperatures and quark densities.
It can be derived via the partition function from the underlying Lagrangian (\ref{eq:Lagrangian}).

Note that $\Phi$ and $\Phib$ are in general complex quantities and \Eq{eq:qqbPotential} contributes an imaginary part to the effective potential which is the manifestation of the fermion Sign Problem in the context of the Polyakov--Quark-Meson model \cite{Fukushima:2006uv,Rossner:2007ik,Mintz:2012mz}. The standard approach to circumvent the Sign Problem of the PQM model is to restrict the Polyakov-loop variables $\Phi$ and $\Phib$ to be two independent, real variables, see e.g.~\cite{Ratti:2005jh}. But by this approach, the state of thermodynamical equilibrium is identified only with a saddle-point but not with a minimum of the effective potential. But only with equilibrium states described by minima of the effective potential one can calculate quasi-equilibrium properties of the system, such as the surface tension and nucleation rate in a first-order phase transition \cite{Mintz:2012mz}.
Another possibility to avoid the Sign Problem is to treat the imaginary part of the effective potential perturbatively which means to simply ignore it in the first order of the approximation \cite{Rossner:2007ik,Mintz:2012mz}. Neglecting the complex part of the effective potential implies that the imaginary part of the Polyakov-loop variables is zero but this approach has the advantage of dealing with the minimization procedure for finding the state of equilibrium.
For the present investigations the results of both approaches are identical.

\section{Results and Discussion}
\label{sec:RD}

As discussed in the previous section, we adjust the parameters in order to reproduce lattice results at zero quark densities \cite{Borsanyi:2010bp,Borsanyi:2010cj,Bazavov:2012bp}, see Fig.~\ref{fig:muI_0}. For a detailed discussion we refer the reader to Refs.~\cite{Haas:2013qwp, Herbst:2013ufa}. It can be seen that our results for the chiral order parameter and thermodynamics either agree quantitatively with lattice results or are at least within the trend of the data. To achieve a better description of the thermodynamics in the phase where chiral symmetry is broken, we augmented the thermodynamics by the contribution of a gas of thermal pions.
The important ingredients to achieve compatibility with the lattice data for the order parameters and thermodynamics are given by the inclusion of the fermionic vacuum loop contribution \eq{eq:QuarkVac}, to unquench the Polyakov-loop potential by applying relation \eq{eq:tYMtglue} and to consider the contribution of a gas of thermal pions.
The inclusion of meson fluctuations in a renormalization group framework would not further improve this agreement, as was shown in Ref.~\cite{Herbst:2013ufa}. The transition region of the chiral and deconfinement observables ($152-164$ MeV) agrees as well with the lattice  \cite{Borsanyi:2010bp}. With this settings we have an adequate framework to investigate the phase structure of strongly interacting matter at nonzero quark and isospin densities and to test the applicability of the model at nonzero densities. The only restriction, which limits the upper value of isospin chemical potential to be considered is given by the omission of the pion condensation, an issue we will address in a future publication \cite{next}.

\begin{figure*}
 	\includegraphics[width=.327\textwidth]{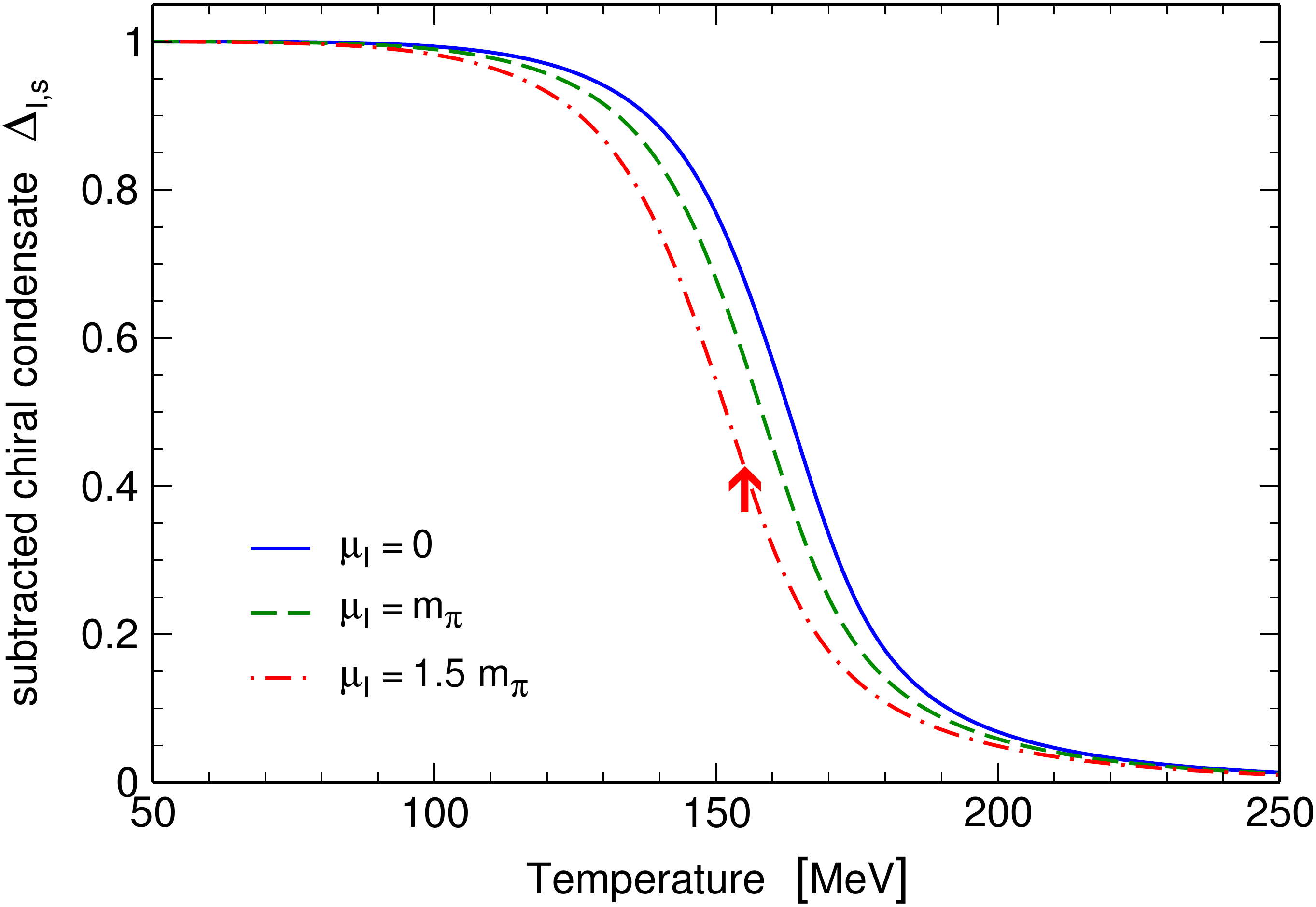}
	\hfill
	\includegraphics[width=.327\textwidth]{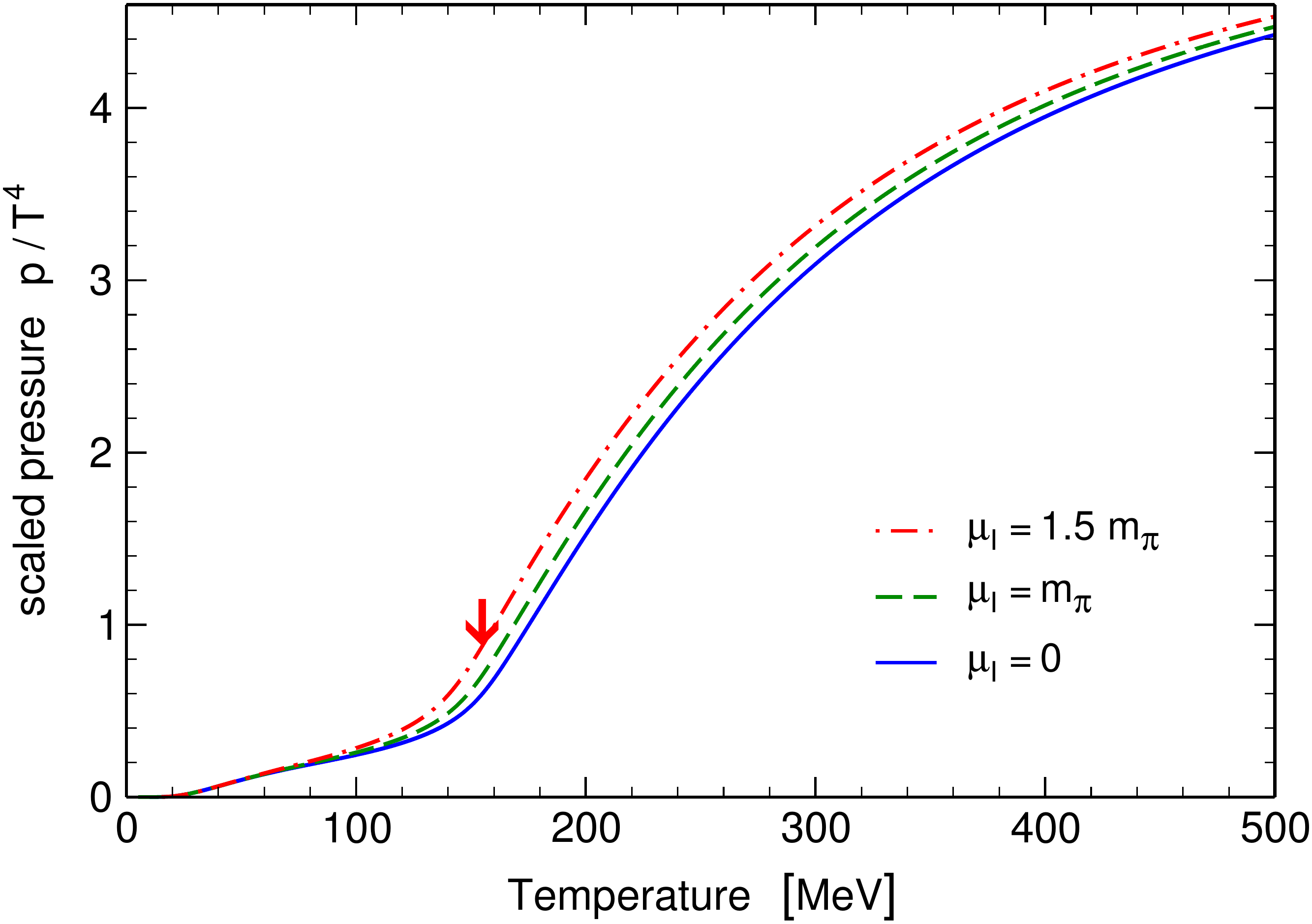}
	\hfill
	\includegraphics[width=.327\textwidth]{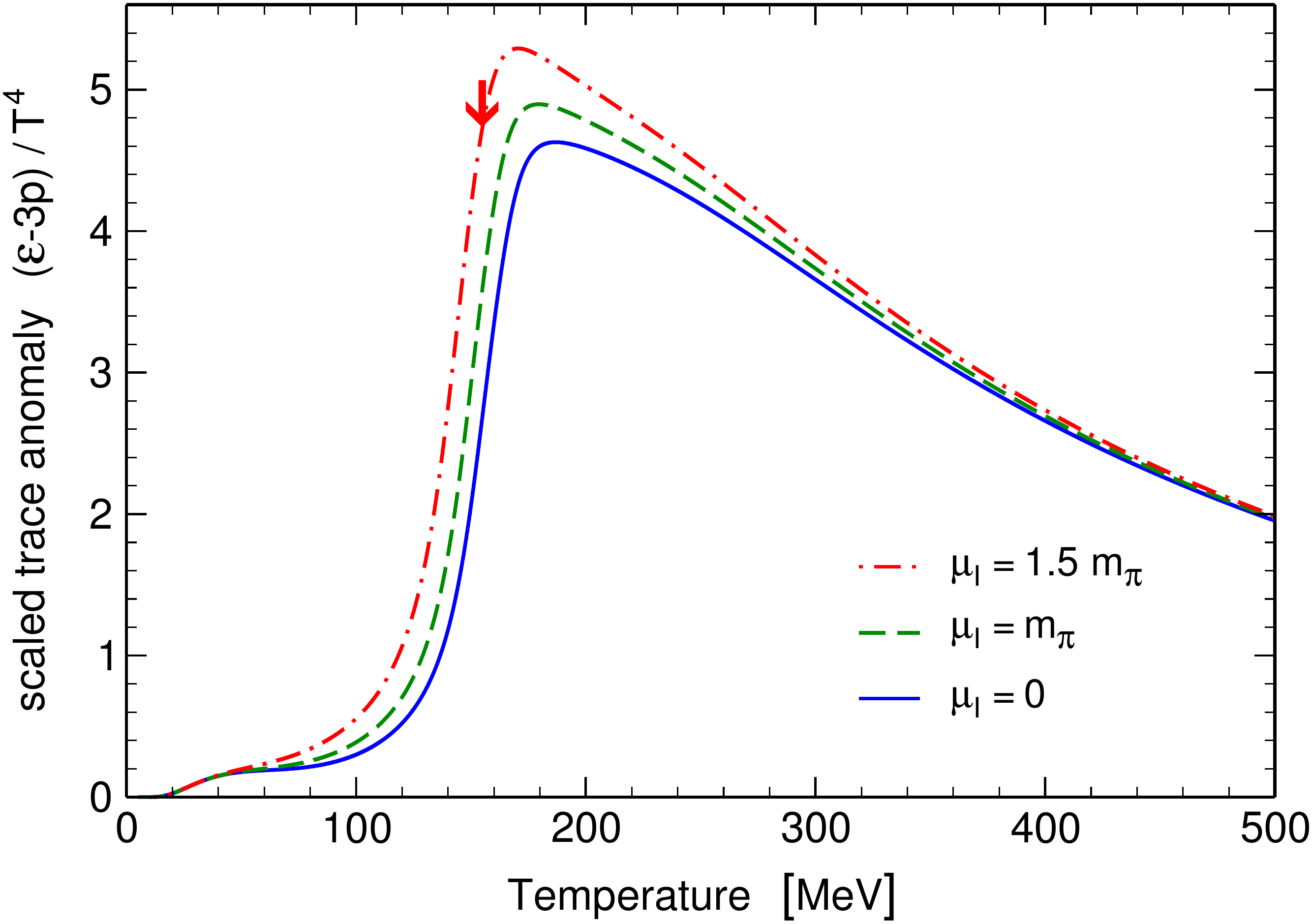}
	\caption{Subtracted chiral condensate, normalized pressure and trace anomaly as functions of the temperature for different isospin at vanishing net quark density. The arrow indicates from which temperature on the in-medium pion mass exceeds the isospin chemical potential in the calculation with $\mu_I=1.5m_\pi$ where $m_\pi=138\,\mrm{MeV}$ is the vacuum pion mass.}
	\label{fig:muIs}
\end{figure*}
\begin{figure*}
 	\includegraphics[width=.48\textwidth]{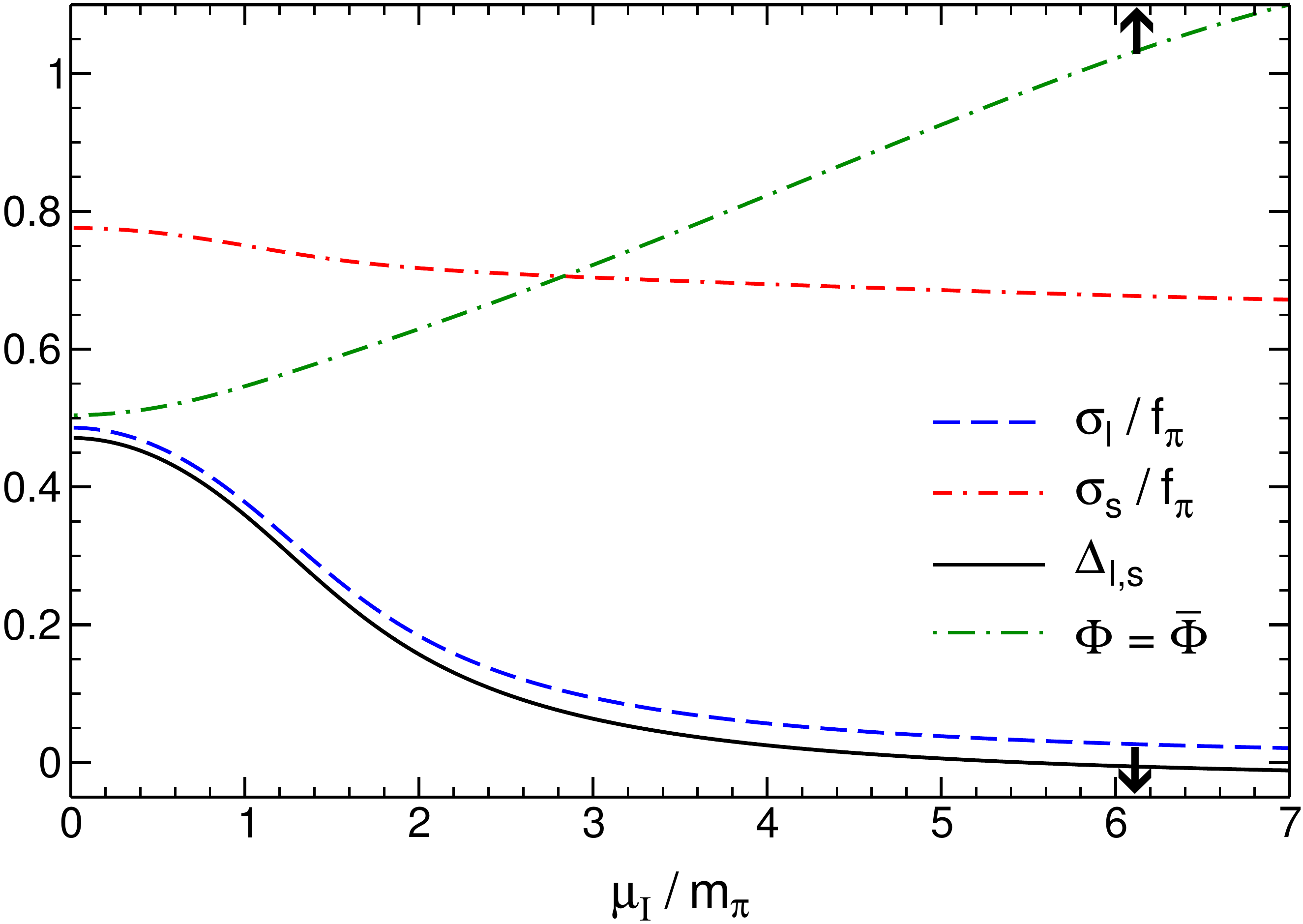}
	\hfill
	\includegraphics[width=.48\textwidth]{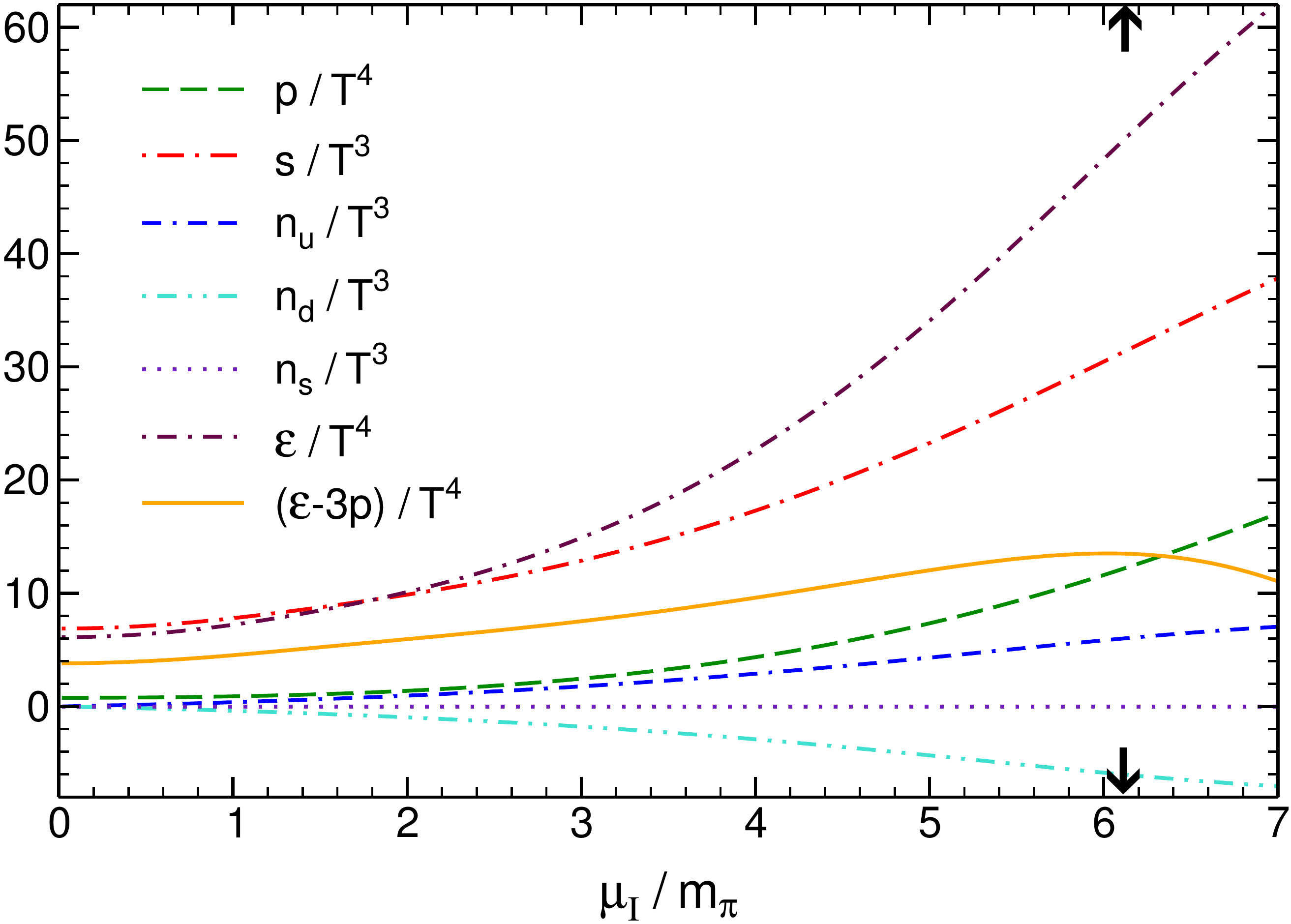}
	\caption{Order parameters (left) and thermodynamic quantities (right) as functions of the isospin at $T=T_{c}^0=164\,\mrm{MeV}$ and vanishing quark density. The arrows indicate from where on the isospin chemical potential exceeds the in-medium pion mass.}
	\label{fig:muI_Tc}
\end{figure*}

To generalize the description to the case of $2+1$ quark flavors, we have to choose the quark chemical potential of strange quarks as well. In the case of supernovae and in the early universe there exists a local $\beta$-equilibrium with respect to weak flavor-mixing interactions so that $\mu_s=\mu_d$. In heavy ion collisions no net strangeness can be produced and the strange quark chemical potential has to be adjusted accordingly. However, the case that can be considered in lattice calculations avoiding the Sign Problem corresponds to the choice $\mu_s$ = 0. Since we take lattice data as a benchmark to compare to, we show results only for the case with vanishing strange chemical potential. For the presented results at vanishing quark density, $\mu_s$ = 0 corresponds to the heavy ion case of zero net strangeness. We discuss briefly the difference to $\mu_s=\mu_d$, and postpone a thorough analysis of this case to a longer publication \cite{next}.

Figure \ref{fig:muIs} shows the impact of moderate isospin on the temperature dependence of the order parameters and thermodynamics at vanishing quark chemical potential. One can see from the plot that a nonzero isospin brings the transition to smaller temperatures. Furthermore, although the chiral condensate decreases appreciably, the
pressure hardly rises and the maximum of the interaction measure experiences a certain increase.
Qualitatively, this dependence of the chiral condensate and pseudo-critical temperature on the isospin is also seen in the lattice calculation \cite{Kogut:2004zg}.
Furthermore, one observes that the chiral condensate gets smaller with increasing isospin for all temperatures. This means a shrinking of the \lq chiral circle{\rq} of the tilted Mexican hat potential with increasing isospin due to an increasing contribution of the thermal quark fluctuations \eq{eq:qqbPotential}. 
At an isospin chemical potential beyond the onset of pion condensation an additional effect would be a rotation from the chiral condensate to the pion condensate \cite{Andersen:2007qv}.

This is also reflected in Fig.~\ref{fig:muI_Tc} where we show the evolution of the order parameters (left) and all relevant thermodynamic quantities (right) as functions of the isospin chemical potential for vanishing quark chemical potential and at the critical temperature $T=T_c^0$ ($=164\,\mrm{MeV}$). One can see that the chiral condensate of the light quarks decreases steeply for increasing isospin chemical potential. On the other hand, chiral symmetry tends to remain broken in the strange quark sector, since the strange quark chemical potential is zero. Only the nonzero temperature reduces the strange chiral condensate and the coupling to the light quarks induces a slight decrease of this condensate for increasing isospin chemical potential.
The Polyakov-loop observables $\Phi$ and $\Phib$ coincide at vanishing quark chemical potential. The different impact of a finite chemical potential of the up quarks $\mu_u$ onto $\Phi$ and $\Phib$ in \Eq{eq:qqbPotential} is just the opposite of the effect of the down quarks with $\mu_d=-\mu_u$. This restriction onto $\Phi=\Phib$ at nonzero isospin but vanishing quark density implies that the Sign Problem is not present in this case.
In the right part of Fig.~\ref{fig:muI_Tc} we see that the pressure and the trace anomaly change moderately with isospin at vanishing baryon chemical potential but the energy density is more sensitive to the variation in isospin.

\begin{figure}
	\includegraphics[width=.48\textwidth]{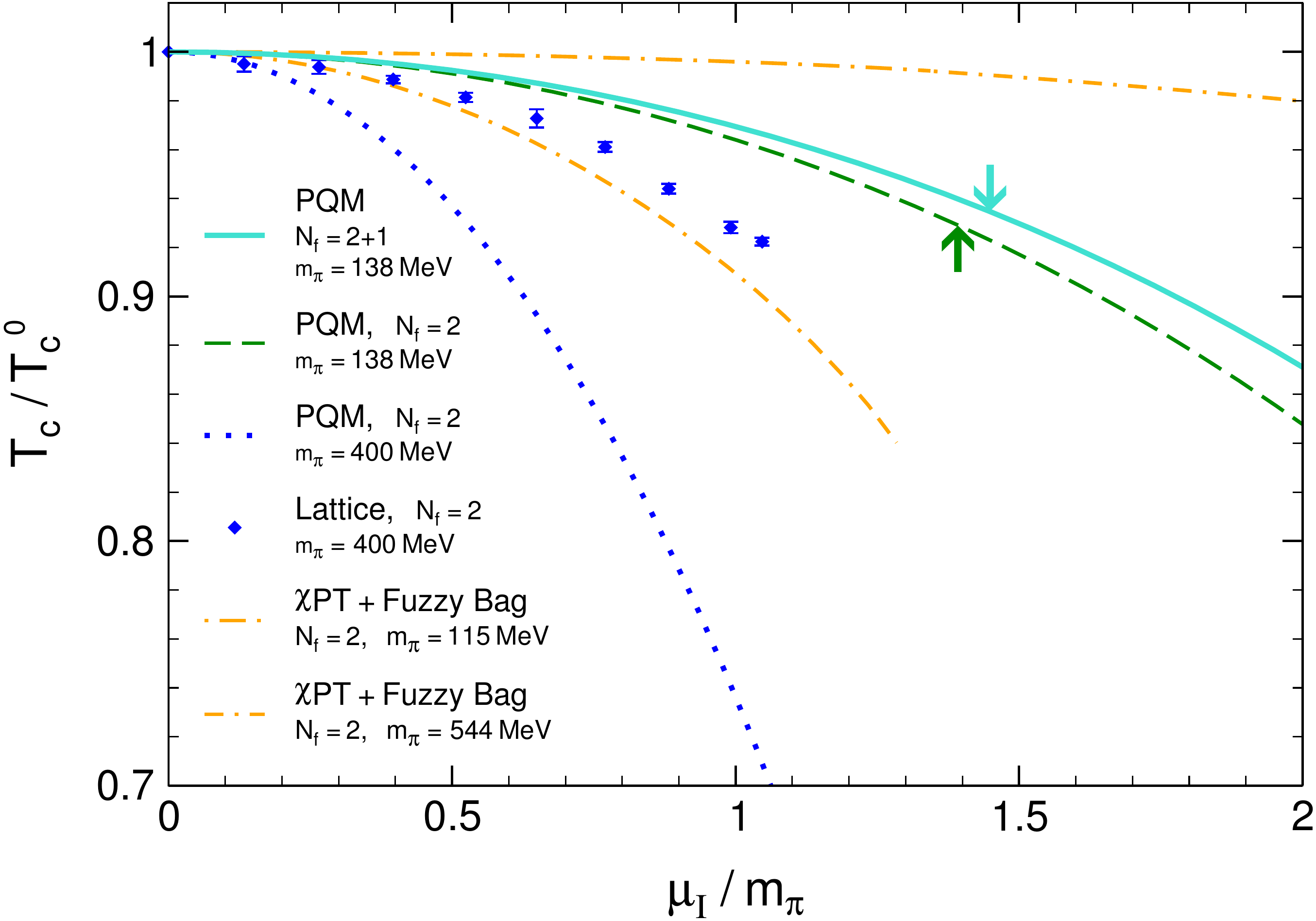}
	\caption{Phase diagram at vanishing net quark density and nonzero isospin. We compare our results to the lattice calculation of Ref.~\cite{Cea:2012ev} and the model calculation of Ref.~\cite{Fraga:2008be}.}
	\label{fig:muI_pd}
\end{figure}

The dependence of the phase structure on the isospin chemical potential is shown in the phase diagram of Fig.~\ref{fig:muI_pd}. We show the decrease of the pseudo-critical temperature with the isospin chemical potential at vanishing baryon chemical potential. We confront our results to the recent lattice simulation of Ref.~\cite{Cea:2012ev} and to results of Ref.~\cite{Fraga:2008be} where the authors combine chiral perturbation theory, including nucleons, for the low-energy sector with the phenomenological fuzzy bag model at high energy.
Note that in all model calculations the pseudo-critical temperatures are linear functions when plotted against $\mu_I^2$ in agreement with lattice data \cite{Kogut:2004zg,Cea:2012ev}.

All these calculations are performed with different pion masses. To correct for the effect of the pion mass onto the abscissae of the temperature-isospin phase diagram, we normalize the isospin chemical potential by the individual pion masses.
So pion condensation at zero temperature where no medium modifications are present sets in at the same point $\mu_I=m_\pi$ in all calculations.
This removes a general offset between calculations with different pion masses. One observes that in the range before the onset of pion condensation the curvature of the transition line is much more sensitive to the pion mass than to the number of quark flavors.

The arrows in Fig.~\ref{fig:muI_pd} indicate the points on the phase transition lines where the isospin chemical potential exceeds the in-medium pion mass, so that condensation of charged pions should no longer be neglected. The upper line of Ref.~\cite{Fraga:2008be} includes a range where a superfluid pion condensate is present (short dot-dashed line), but the curvature of the transition line is hardly affected by this correction. The lattice calculation, on the other hand, is performed for values of the isospin chemical potential below the onset of pion condensation.

We show results for $2+1$ and 2 flavors and physical pion mass (light blue, solid curve and green, dashed curve) and for $2$ flavors and $m_{\pi}=400$ MeV (dark blue, dotted curve), the latter being more appropriate to compare to the currently available lattice data \cite{Cea:2012ev}. In the case with $m_\pi=400$\,MeV, we considered the dependence of the nucleon mass on the pion mass according to Ref.~\cite{Bali:2012qs} and increased the scalar coupling accordingly. The coupling $\lambda$ is also adjusted to preserve the mass difference $m_{\sigma}^{2}-m_{\pi}^{2}$ in the vacuum.

The results for $2$ flavors and $m_{\pi}=400$ MeV (dark blue, dotted line in Fig.~\ref{fig:muI_pd}) show a decrease with $\mu_I$ that is significantly larger than the one obtained on the lattice. If we included meson fluctuations within the RG-improved Quark-Meson model \cite{Herbst:2010rf,Skokov:2010wb} we would expect a decrease in the curvature of the phase transition line. So our result would be in line with the observation that the pseudo-critical temperature drops faster with increasing baryon chemical potential in low-energy effective models compared to functional methods \cite{Fischer:2013eca} and the lattice \cite{Endrodi:2011gv}.\\
In Ref.~\cite{Bratovic:2012qs} it was shown that including repulsive vector interactions to a PNJL study can lead to a slope of the phase transition line at nonzero baryon chemical potential that corresponds to the data of lattice calculations. But one should be aware that by including the vector meson exchange to such a model this fails to describe lattice results of quark number susceptibilities as was shown in Ref.~\cite{Steinheimer:2010sp}.\\
Some studies compared the curvature of the crossover line along the isospin and light quark chemical potential axes within their calculations. While in older investigations the pseudo-critical temperatures are almost identical along both axes \cite{Toublan:2003tt,Toublan:2004ks,Kogut:2004zg}, Ref.~\cite{Cea:2012ev} found a difference of about 10\,\% of the slope parameters. The framework applied by us implies the former result but we plan to address this issue in a future publication \cite{next} when we will consider thermal fluctuations of the pion fields.

Our results and those of Ref.~\cite{Fraga:2008be} at two different pion masses illustrate the impact of the pion mass on the phase transition line. A larger pion mass increases the curvature of the transition line significantly.
Hence, our calculation provides a lower limit of the pseudo-critical temperature at nonzero isospin chemical potential for the case of $2+1$ quark flavors and physical quark masses. Results from the lattice in the physical limit would be very helpful to constrain the applicability of effective chiral models at nonzero densities.

The impact of the pion mass on the transition line is much larger than that of adding strange quarks to the system. Our results for $N_f=2$ lies relatively close but below the $N_f=2+1$ result and the transition temperatures only deviate by two percent just before the onset of pion condensation.

An effect of similar magnitude and in the same direction has a non-vanishing strange quark density with $\mu_s=\mu_d$, as is the case in the astrophysical and cosmological scenarios. It increases the curvature of the transition line. 
For instance, the value of the transition temperature in the region just before the onset of pion condensation becomes two percent smaller.

In a future publication \cite{next}, we shall consider nonzero pion fields to include the possibility of pion condensation at larger isospin, as well as meson fluctuations included by a renormalization group improvement. Finally, this framework can be extended to address the cosmological QCD phase transition at non-vanishing lepton number.

\begin{acknowledgments}
ESF is grateful to Massimo D'Elia for his kind hospitality during a visit to the University of Pisa and for various fruitful discussions. 
The authors also thank him for providing the lattice data. We thank Let\'icia Palhares for providing the data of Ref.~\cite{Fraga:2008be} and for discussions. We are also grateful to Nils Strodthoff for discussions.
This work is supported by BMBF under grant 05P12VHCTG.
The work of ESF was financially supported by the Helmholtz International Center for FAIR within the framework of the LOEWE program (Landesoffensive zur Entwicklung Wissenschaftlich-\"Okonomischer Exzellenz) launched by the State of Hesse.
RS acknowledges support by the Heidelberg Graduate School HGSFP, the Helmholtz Graduate School HGS-HIRe and the Graduate Program for Hadron and Ion Research through the GSI Helmholtzzentrum and the Frankfurt Institute for Advanced Studies.
\end{acknowledgments}

\bibliography{references}

\end{document}